# Target Trial Emulation with the R Package TTE: A Tutorial and Methodological Guide

Hisashi Noma

*Department of Interdisciplinary Statistical Mathematics, The Institute of Statistical Mathematics, Tokyo, Japan*
*Department of Statistical Science, The Graduate University for Advanced Studies, SOKENDAI, Tokyo, Japan*

Correspondence: Hisashi Noma, 10-3 Midori-cho, Tachikawa, Tokyo 190-8562, Japan
E-mail: noma@ism.ac.jp | ORCID: 0000-0002-2520-9949

## SUMMARY

Target trial emulation structures observational causal analyses around the protocol of an ideal randomized trial. By aligning eligibility, treatment assignment, time zero, and follow-up, it can reduce avoidable biases, but implementation still requires coordinated decisions about data construction, inverse probability weighting, diagnostics, outcome models, standardization, competing risks, and uncertainty estimation. This article provides a self-contained methodological guide and practical tutorial for TTE, an R package for target trial emulations with longitudinal observational data. We describe target trial protocols, intention-to-treat and per-protocol estimands, identification assumptions, baseline and person-period data structures, temporal ordering for longitudinal weights, stabilized treatment and censoring weights, weight truncation, balance and effective-sample-size diagnostics, weighted pooled discrete-time survival models, model-based standardization, competing-risk analysis, weighted Kaplan-Meier and Aalen-Johansen estimation, and cluster bootstrap at the original-individual level. Two fully synthetic examples illustrate end-to-end workflows: sodium-glucose cotransporter 2 inhibitor versus dipeptidyl peptidase-4 inhibitor initiation with all-cause death, and sequentially nested angiotensin receptor blocker versus calcium channel blocker trials with heart-failure hospitalization and competing death. The examples show how to obtain and diagnose estimates in R and interpret relative hazards, absolute risks, cumulative incidence, and differences between intention-to-treat and per-protocol effects.

## 1. INTRODUCTION

Randomized trials define a treatment question through a protocol: who is eligible, which strategies are compared, when treatment is assigned, when follow-up begins, which outcomes are measured, and which causal contrast is targeted. Observational studies often begin instead with an available database and a broad comparison of exposed and unexposed people. When eligibility, treatment assessment, and the start of follow-up are not aligned, apparently sophisticated analyses can inherit immortal-time bias, selection bias, or ambiguity about the intervention being compared. Target trial emulation reverses that order. The investigator first specifies the randomized trial that would answer the question and then operationalizes each protocol component using observational data (Hernán & Robins, 2016; Hernán et al., 2022, 2025).

The framework is particularly useful for longitudinal healthcare databases, in which treatment decisions, eligibility, measurements, switching, discontinuation, loss to follow-up, and competing events unfold over time. An active-comparator new-user design can make treatment groups more clinically comparable and avoid conditioning on survival and tolerance under prior therapy (Ray, 2003). Repeated eligibility can be used to emulate a sequence of nested trials, allowing an individual to contribute at multiple treatment-decision times while preserving the original person as the sampling unit (Danaei et al., 2013). Per-protocol effects can be studied by artificially censoring follow-up at strategy deviation and weighting for the resulting time-dependent selection. These design choices are not interchangeable technical options: they define different interventions and estimands.

The statistical ingredients of a target trial emulation are individually familiar. They include propensity scores and inverse probability weights, marginal structural models, grouped or discrete-time survival models, standardization, competing-risk methods, and cluster-robust or bootstrap inference (Robins et al., 2000; Cole & Hernán, 2008; Prentice & Gloeckler, 1978; Andersen et al., 2012). Their joint implementation is nevertheless demanding. Analysts must preserve the temporal sequence of covariates, treatment, censoring, and outcomes; accumulate longitudinal weights within the correct person-trial; avoid using post-outcome information; assess whether weighting has created an empirically supported pseudo-population; and translate relative model coefficients into clinically interpretable absolute risks.

The R package TTE (Noma, 2026) was developed to make this workflow explicit and inspectable. Rather than placing the full analysis inside one opaque procedure, the package provides separate functions for data checks, sequential-trial expansion, treatment and censoring weights, weight combination, balance and support diagnostics, weighted discrete-time outcome models, standardization, weighted survival and cumulative-incidence curves, and individual-level bootstrap. Intermediate predicted probabilities, interval factors, cumulative weights, fitted models, and risk-set summaries remain accessible. This modularity supports practical analysis, collaborative review, and reproducible reporting.

This article is written for investigators who understand ordinary R data frames, formulas, and regression models and who need a coherent route from a target trial protocol to an interpretable analysis. The article is also intended to be readable by graduate students learning target trial emulation, but the workflow is not limited to demonstration: the same functions and diagnostics are designed for real-world comparative-effectiveness analyses.

This article documents an original open-source implementation that integrates the core design and analysis components of target trial emulation into a modular, inspectable, and reproducible R workflow. Its contribution is computational and integrative rather than a new identification theory or causal estimator. The implementation is intended to support both applied comparative-effectiveness analyses and methodological training. A valid analysis still depends on a scientifically defensible target trial, credible causal assumptions, appropriate measurement, and sensitivity analyses. The TARGET Statement should be used when reporting an observational study that explicitly emulates a target trial (Cashin et al., 2025).

## 2. THE TARGET TRIAL MUST BE SPECIFIED BEFORE THE SOFTWARE IS USED

### *2.1. Protocol components and alignment*

A target trial protocol defines the causal question before estimation begins. Table 1 summarizes the principal components and the corresponding responsibility of the analyst. TTE begins after these components have been specified and operationalized; it cannot infer a clinically meaningful intervention from the data.

**Table 1.** Core components of a target trial protocol and their implications for implementation

| Component | Question to answer before analysis | Frequent implementation problem |
|---|---|---|
| Objective | What causal question would the ideal randomized trial answer? | Starting from available variables rather than a defined decision problem |
| Eligibility | Who could enter the trial at each treatment-decision time? | Using information measured after time zero or applying asymmetric criteria |
| Treatment strategies | What precisely distinguishes strategy 0 from strategy 1? | Treating heterogeneous versions, dose changes, or add-on therapy as one intervention without justification |
| Assignment | What observational adjustment is intended to emulate randomization? | Choosing covariates only for prediction, or adjusting for colliders or post-treatment variables |
| Time zero | When do eligibility, strategy assignment, and follow-up begin? | Misalignment that induces immortal time or selection bias |
| Follow-up and censoring | Which events end observation, and which create artificial censoring? | Confusing loss to follow-up, competing events, strategy deviation, and administrative end |
| Outcomes | What event definition, induction period, and competing events apply? | Outcome ascertainment that depends differentially on strategy or occurs before a biologically plausible risk window |
| Causal contrast | Is the target an initiation effect, a sustained-strategy effect, or another contrast? | Reporting an analysis whose censoring rule does not correspond to the stated estimand |
| Analysis plan | How will confounding, censoring, absolute risk, competing events, and uncertainty be handled? | Selecting models after viewing treatment-effect results without prespecified diagnostics or sensitivity analyses |

The central alignment principle is that eligibility, treatment assignment, and time zero refer to the same treatment-decision occasion. In a new-user study, time zero is usually the first qualifying prescription after a washout period. In a sequential-trial design, a new trial entry is created whenever a person is eligible at a prespecified decision time. Follow-up for each entry begins at that trial-specific time zero.

An induction period may be used when events immediately after treatment assignment cannot plausibly be caused by the strategy or when the database records exposure with a short delay. An induction period changes the operational target trial and should not be inserted merely because early results are inconvenient. A grace period is different: it defines a window during which a strategy may be initiated or complied with. Complex grace-period strategies often require cloning and censoring methods beyond the standard static-strategy workflow covered here. In TTE, the induction argument specifies the number of initial follow-up intervals excluded from outcome ascertainment; it does not redefine the treatment assignment or the trial-specific time zero.

### *2.2. Intention-to-treat and per-protocol estimands*

An intention-to-treat (ITT) estimand compares treatment initiation strategies according to the baseline assignment, regardless of later switching, discontinuation, or add-on therapy. In an observational emulation, the label "ITT" is an analogy: treatment is not randomized, and baseline exchangeability must be created by design and adjustment. The estimand is often clinically relevant because it approximates the effect of choosing one initial strategy rather than another in routine practice.

A per-protocol (PP) estimand compares sustained adherence to the strategies defined in the protocol. Follow-up is artificially censored when a person deviates from the assigned strategy. Because deviation may depend on prognosis and time-varying patient characteristics, simply censoring at deviation generally creates selection bias. Inverse probability of censoring weights are used to reweight those who remain compatible with the strategy so that they represent people with similar measured histories who deviated.

The two estimands answer different questions. A PP estimate need not be closer to the truth than an ITT estimate, and numerical differences between them are not by themselves evidence of bias. PP estimation usually requires stronger data and assumptions, especially sufficiently frequent measurement of time-varying predictors of adherence and outcome. The protocol must state exactly what counts as discontinuation, switching, add-on treatment, dose deviation, or a clinically allowed treatment modification.

### *2.3. Identification assumptions*

The software returns numerical estimates, but causal interpretation requires assumptions that cannot be established by a package (Hernán & Robins, 2020).

Consistency requires that the observed outcome under the received strategy corresponds to the potential outcome under the strategy as defined. This is difficult when treatment versions are vague or clinically heterogeneous. Conditional exchangeability requires no unmeasured confounding of treatment and outcome at baseline and, for PP analyses, no unmeasured time-varying predictors of artificial censoring and outcome. Positivity requires a nonzero probability of each strategy, and of remaining uncensored, within relevant covariate histories. No interference requires that one person's strategy does not alter another person's outcome, unless the estimand explicitly incorporates such effects. Correct temporal measurement, missing-data handling, and adequate nuisance and outcome models are also needed.

Balance diagnostics can show whether measured baseline covariates are similar after weighting. They cannot show that exchangeability holds for unmeasured factors. Likewise, finite and apparently well-behaved weights do not prove positivity; they provide empirical information about support in the observed sample.

### *2.4. Design patterns covered by TTE*

TTE is most directly suited to common static-strategy analyses. A single active-comparator new-user trial has one baseline row per individual and one sequence of follow-up intervals. Sequentially nested trials allow repeated trial entries when eligibility recurs. Baseline treatment weights address nonrandom assignment; longitudinal weights address loss to follow-up or artificial censoring; weighted discrete-time models estimate relative event-rate contrasts; and standardization or weighted risk-set estimators produce absolute risks.

Competing events require an explicit estimand. For a nonfatal outcome such as heart-failure hospitalization, death prevents subsequent observation of the event. A cause-specific hazard model describes the instantaneous event rate among people still event-free, whereas the cumulative incidence function describes the probability of experiencing the event by a horizon in the presence of death. These quantities are related but not interchangeable. TTE estimates cumulative incidence from separate cause-specific models or directly by a weighted Aalen-Johansen estimator.

## 3. DATA STRUCTURES AND THE MODULAR TTE WORKFLOW

### *3.1. Baseline and person-period data*

TTE distinguishes two core data structures. A baseline dataset contains one row per trial entry. In a simple new-user study, this is usually one row per individual. In sequentially nested trials, the same individual can contribute several baseline rows because eligibility can be met at several calendar times.

A person-period dataset contains one row for each follow-up interval in which a person-trial remains at risk. Typical variables include the original individual identifier, trial identifier, zero-based interval index, baseline treatment, interval-specific outcomes, loss-to-follow-up and adherence indicators, baseline and time-varying covariates, and analysis weights. The format is the same one used for pooled discrete-time survival regression.

The independent sampling unit is generally the original individual, not the person-period row and not the trial entry. When one individual contributes many intervals or several nested trials, covariance estimation and bootstrap must preserve all those records together. TTE therefore distinguishes the original id from the trial identifier.

### *3.2. Structural checks and sequential-trial construction*

check_tte() checks whether a person-period dataset is structurally compatible with the intended analysis. It identifies duplicated individual-trial-time combinations, noninteger or gapped interval sequences, records after a terminal event, multiple terminal events, within-trial changes in a baseline treatment variable, and missing, nonfinite, or negative weights. These checks detect data errors; they do not assess causal validity.

seqdesign_tte() creates a new trial entry at each eligible treatment-decision time and expands that entry into follow-up records. The user supplies the original individual identifier, ordered calendar time, eligibility indicator, observed treatment, outcome, optional censoring indicator, maximum follow-up, and induction period. The function does not decide whether the washout period, eligibility rule, or strategy definition is scientifically appropriate. Those rules must already be encoded in the source data.

### *3.3. Modular analysis stages*

Figure 1 shows the intended workflow. Protocol operationalization precedes the software. Weighting, diagnostics, outcome modeling, and marginal-effect estimation are separate stages. boot_tte() crosses these stages because complete uncertainty propagation requires repeating the entire pipeline in each resample.

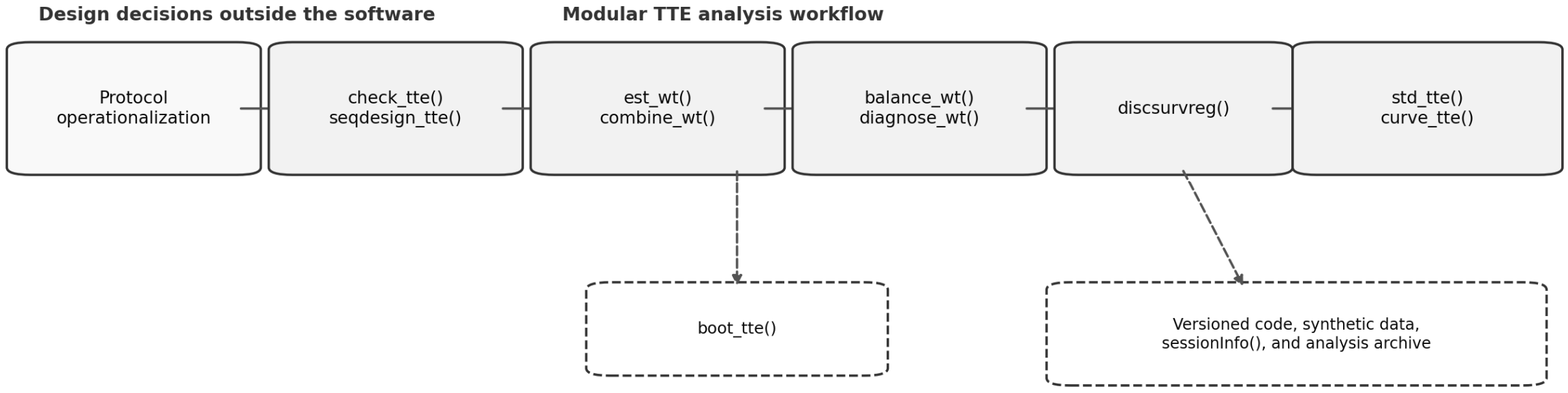

**Figure 1.** Modular workflow implemented in TTE. Protocol definition and causal identification assumptions precede the software. Dashed boxes indicate cross-cutting uncertainty and reproducibility components.

Table 2 maps the main analysis questions to package functions and outputs.

**Table 2.** Principal TTE functions and their role in the workflow

| Stage | Function | Main output |
|---|---|---|
| Structural checking | check_tte() | Duplicate, timing, post-event, treatment, and weight diagnostics |
| Sequential-trial construction | seqdesign_tte() | Stacked person-trial follow-up data |
| Treatment, censoring, or adherence weights | est_wt() | Fitted probability models, observed probabilities, interval factors, cumulative and final weights |
| Weight combination | combine_wt() | Product weight, truncation limits, normalized or unnormalized analysis weight |
| Baseline balance | balance_wt() | Unweighted and weighted means, standard deviations, and standardized mean differences |
| Weight and support diagnostics | diagnose_wt() | Weight distribution, effective sample size, group summaries, and effective risk sets |
| Weighted outcome model | discsurvreg() | Pooled discrete-time model with individual-cluster covariance |
| Model-based marginal effects | std_tte() | Survival, risk, cumulative incidence, risk difference, risk ratio, and NNT or NNH |
| Direct weighted curves | curve_tte() | Weighted Kaplan-Meier or Aalen-Johansen curves |
| Complete resampling | boot_tte() | Original-individual bootstrap replicates and percentile intervals |

TTE uses ordinary R formulas and data frames within the R environment (R Core Team, 2026). Weight arguments can be numeric vectors, column names, or objects returned by est_wt() and combine_wt(). S3 methods such as print(), summary(), plot(), weights(), coef(), vcov(), confint(), predict(), and as.data.frame() are available as appropriate. Intermediate quantities are retained rather than discarded after a final estimate is produced.

### *3.4. Follow-up-time convention*

The internal person-period variable is a zero-based interval index. Thus, time = 0 denotes the first interval, which ends at elapsed follow-up time 1. TTE stores and reports public summaries on the elapsed-time scale. Consequently, times = 0:59 represents 60 intervals; summary(x, horizon = 60) for a standardized object and summary(x, time = 60) for a weighted curve both refer to the end of 60 follow-up intervals.

This convention should be stated explicitly in an applied analysis. Confusing an interval index with elapsed time can shift reported horizons by one interval and can also misalign longitudinal weights.

## 4. STATISTICAL METHODS IMPLEMENTED IN TTE

### *4.1. Baseline treatment weights*

Let $A_i$ denote baseline treatment for trial entry $i$, $a_i$ the observed strategy, and $L_i$ measured baseline covariates. The default stabilized inverse probability of treatment weight for a binary or multinomial strategy is

$$SW_i^A = \frac{\Pr(A_i = a_i)}{\Pr(A_i = a_i \mid L_i)}. \tag{1}$$

est_wt(type = "treatment") fits a binary logistic model when treatment has two levels and a multinomial model when it has more than two. When stabilize = TRUE and no numerator formula is supplied, the numerator is the empirical marginal probability of the observed treatment category. When stabilize = FALSE, the numerator is one. Predicted probabilities are bounded away from zero and one by eps to avoid undefined numerical weights. Probability clipping is a numerical safeguard, not a remedy for empirical positivity violations; analysts should report the frequency of clipping when it is non-negligible.

Covariate selection remains a scientific decision. Variables should be chosen because they are needed for conditional exchangeability, not merely because they improve treatment prediction. Strong instruments can increase variance, and colliders can introduce bias (Brookhart et al., 2006). Flexible functional forms may be needed for continuous covariates and interactions that determine treatment choice.

### *4.2. Longitudinal censoring and adherence weights*

Let $R_{ik} = 1$ indicate that person-trial $i$ remains observed or strategy-compatible through interval $k$. A stabilized cumulative continuation weight through the start of interval $t$ can be written as

$$SW_{it}^R = \prod_{k=0}^{t-1} \frac{\Pr\left(R_{ik} = 1 \mid \bar{R}_{i,k-1} = \mathbf{1}, \bar{H}_{ik}^{\text{red}}\right)}{\Pr\left(R_{ik} = 1 \mid \bar{R}_{i,k-1} = \mathbf{1}, \bar{H}_{ik}^{\text{full}}\right)}. \tag{2}$$

Conditioning on the previous continuation history means that each numerator and denominator probability is evaluated among person-trials that remain in the relevant continuation risk set immediately before interval $k$; this condition is satisfied by convention at $k = 0$.

The reduced history in the numerator commonly contains treatment, follow-up time, calendar period, and selected baseline variables; the full history in the denominator additionally contains measured predictors of continuation and outcome. For type = "censoring" or type = "adherence", est_wt() calculates interval-specific ratios and, by default, multiplies them within each person-trial trajectory.

The lag argument aligns the cumulative weight with the outcome interval. With lag = 1, the first outcome interval receives weight 1, and interval $t$ receives the product accumulated through the preceding interval. This is appropriate when stay_ltfu, stay_adherent, or another continuation indicator describes remaining uncensored through the next interval. Figure 2 illustrates the ordering. Applied analyses must verify how each indicator and time-varying covariate is measured in the source data; no universal lag can repair an incorrectly defined longitudinal dataset.

### *4.3. Combined weights, truncation, and normalization*

A common ITT analysis weight combines the baseline treatment weight and a loss-to-follow-up weight:

$$W_{it}^{\text{ITT}} = SW_i^A \, SW_{it}^{\text{LTFU}}. \tag{3}$$

A PP analysis replaces the loss-to-follow-up component by a joint continuation weight that accounts for both observation and strategy adherence:

$$W_{it}^{\text{PP}} = SW_i^A \, SW_{it}^{\text{joint censoring}}. \tag{4}$$

| | 0–1 | 1–2 | 2–3 | 3–4 | 4–5 |
|---|---|---|---|---|---|
| | interval 0 | interval 1 | interval 2 | interval 3 | interval 4 |
| Observed continuation indicator | $R_0$ | $R_1$ | $R_2$ | $R_3$ | $R_4$ |
| Interval factor | $f_0$ | $f_1$ | $f_2$ | $f_3$ | $f_4$ |
| Cumulative factor | $f_0$ | $f_0f_1$ | $f_0f_1f_2$ | $f_0f_1f_2f_3$ | $f_0f_1f_2f_3f_4$ |
| Analysis weight with lag = 1 | 1 | $f_0$ | $f_0f_1$ | $f_0f_1f_2$ | $f_0f_1f_2f_3$ |

The outcome contribution in interval t uses the cumulative weight available at the start of that interval.

**Figure 2.** Temporal alignment of interval-specific and cumulative longitudinal weights. With lag = 1, the contribution in interval $t$ uses the cumulative information available at the start of that interval.

combine_wt() multiplies components of equal length. The product can be truncated at empirical quantiles, with the 1st and 99th percentiles as the default. It can be left unnormalized, normalized to mean one, or normalized so that the sum equals the number of rows. When the scientific analysis regularizes the final product rather than each component, component objects should be created with `truncate = c(0, 1)` and truncation should be applied once in combine_wt(). The default truncation is therefore an active analysis choice rather than a neutral computational setting; it should be explicitly reported, and truncate = c(0, 1) should be used when no truncation is intended.

Truncation trades bias against variance and should be prespecified or examined in sensitivity analyses. A large treatment-effect change under modest alternative truncation limits is a warning that the analysis depends on sparse support. Normalization changes the scale of the weights but not point estimates from many weighted regression models; it can affect descriptive summaries and some variance calculations, so the selected rule should be reported.

### 4.4. Covariate balance, support, and effective sample size

For a continuous covariate $X$, the unweighted standardized mean difference is

$$\mathrm{SMD}(X) = \frac{\bar{X}_1 - \bar{X}_0}{\sqrt{(s_0^2 + s_1^2)/2}}. \tag{5}$$

balance_wt() replaces the group means by weighted means for the weighted SMD and retains the same unweighted pooled standard deviation in the denominator. Factor contrasts and user-specified transformations are expanded through model.matrix(), so nonlinear terms can be examined separately. An absolute SMD below 0.10 is a common diagnostic convention, not a proof that confounding has been removed (Austin & Stuart, 2015).

The effective sample size (ESS) for weights $w_i$ is

$$\mathrm{ESS} = \frac{(\sum_i w_i)^2}{\sum_i w_i^2}. \tag{6}$$

diagnose_wt() reports the mean, standard deviation, selected quantiles, minimum, maximum, treatment-specific summaries, overall ESS, and effective risk-set sizes by treatment and follow-up interval. A nominally large study can contain little effective information late in follow-up if a few highly weighted individuals dominate a risk set. Treatment-specific effective risk sets are therefore more informative than the raw number of person-period rows.

In longitudinal person-period data, an ESS calculated over rows is primarily a diagnostic of weight concentration and should not be interpreted as the number of independent individuals. Treatment-specific effective risk sets at clinically relevant horizons are therefore especially informative.

Diagnostics should be interpreted together. Good baseline SMDs do not compensate for extreme longitudinal weights, and an acceptable overall ESS can conceal near-empty late risk sets. Conversely, some weight variability is expected because the purpose of weighting is to change the empirical distribution.

### 4.5. Weighted pooled discrete-time outcome models

For an interval-specific event indicator $Y_{it}$ and covariates $Z_{it}$, a typical weighted model is

$$\mathrm{cloglog}\{\Pr(Y_{it} = 1 \mid I_{it} = 1, A_i, Z_{it})\} = \alpha(t) + \beta A_i + \gamma^{\mathsf{T}} Z_{it}. \tag{7}$$

Here, $I_{it} = 1$ indicates that person-trial $i$ is alive, uncensored, and free of all event types relevant to the analysis at the start of interval $t$.

discsurvreg() fits this model as a weighted generalized linear model to person-period data. The default is quasibinomial(link = "cloglog"). Under the fitted proportional-hazards form, $\exp(\beta)$ is the ratio of interval-specific integrated hazards under a grouped proportional-hazards interpretation and corresponds to a continuous-time hazard ratio when proportional hazards hold within intervals. Follow-up time may be represented by indicators, polynomials, or regression splines. A treatment-by-time interaction can be added when proportional hazards are not assumed, but the resulting treatment effect must then be summarized as a time-varying contrast or through standardized risks rather than a single coefficient.

Covariance estimation is clustered by the original individual. The standard option uses an HC0 cluster sandwich covariance through sandwich::vcovCL() (Zeileis et al., 2020). The optional Morel-Bokossa-Neerchal correction is

$$V_{\mathrm{MBN}} = q_1 q_2 V_{\mathrm{HC0}} + \delta\gamma V_{\mathrm{model}}. \tag{8}$$

Here $q_1 = (L-1)/(L-p)$, $q_2 = K/(K-1)$, $q_3 = \mathrm{tr}(V_{\mathrm{HC0}}A)/p$, $\delta = \min\{0.5, p/(K-p)\}$, $\gamma = \max(1, q_3)$, $L$ is the number of person-period rows, $K$ the number of independent individuals, $p$ the number of coefficients, $V_{\mathrm{model}}$ the model-based covariance, and $A = V_{\mathrm{model}}^{-1}$. The correction requires $K > p$ (Morel et al., 2003). It addresses finite-sample covariance estimation, not confounding, positivity, or model misspecification.

### *4.6. Standardization to marginal survival and risk*

A hazard ratio is a relative rate contrast and does not directly state how many events would occur by a clinically meaningful horizon. std_tte() creates a counterfactual copy of every member of a target population under each treatment value, predicts interval-specific event probabilities, recursively accumulates survival and risk, and averages the individual predictions.

Let $p_{ij}(a)$ be the predicted event probability for individual $i$ in interval $j$ if treatment were set to $a$. Without a competing event, survival is updated by

$$S_{ij}(a) = S_{i,j-1}(a)\{1 - p_{ij}(a)\}, \tag{9}$$

and cumulative risk by

$$F_{ij}(a) = F_{i,j-1}(a) + S_{i,j-1}(a)p_{ij}(a). \tag{10}$$

The individual curves are averaged equally or with optional nonnegative target_weights. For the first two treatment strategies, TTE reports strategy-specific survival and risk, risk differences, risk ratios, survival differences, and time-specific numbers needed to treat or harm. The target population used for averaging is part of the estimand and should be reported. Standardizing to the original eligible cohort, to a subgroup, or to an external target population answers different questions.

Standardization is model based. It can smooth sparse risk sets and adjust residual measured imbalance if covariates are included in the outcome model, but it depends on the outcome-model functional form. Direct weighted curves provide a useful complementary analysis.

Standardization is directly applicable when the fitted model contains treatment, functions of follow-up time, baseline covariates, and other variables whose counterfactual values are defined in the target dataset. Treatment-dependent time-varying covariates require separate models for their counterfactual trajectories and are not generated automatically by std_tte(). Including covariates in a weighted outcome model does not, by itself, make the estimator doubly robust.

### *4.7. Standardized cumulative incidence with competing risks*

When a competing event precludes the event of interest, one minus Kaplan-Meier generally overstates the cumulative incidence of the event of interest (Andersen et al., 2012). TTE fits separate cause-specific discrete-time models and combines their predicted probabilities during standardization.

Let $p_{1ij}(a)$ and $p_{2ij}(a)$ denote predictions from the two separately fitted complementary-log-log models. TTE maps each prediction to a latent integrated hazard contribution by

$$h_{kij}(a) = -\log\{1 - p_{kij}(a)\}, \qquad k = 1,2. \tag{11}$$

With $h_{Tij}(a) = h_{1ij}(a) + h_{2ij}(a)$, the total probability of any event in the interval is allocated to each cause in proportion to its integrated hazard:

$$q_{1ij}(a) = \{1 - \exp[-h_{Tij}(a)]\}\frac{h_{1ij}(a)}{h_{Tij}(a)}, \tag{12}$$

$$q_{2ij}(a) = \{1 - \exp[-h_{Tij}(a)]\}\frac{h_{2ij}(a)}{h_{Tij}(a)}. \tag{13}$$

If $h_{Tij}(a) = 0$, both increments are zero. The cause-specific cumulative incidences and event-free survival are updated by

$$F_{1ij}(a) = F_{1i,j-1}(a) + S_{i,j-1}(a)q_{1ij}(a), \tag{14a}$$

$$F_{2ij}(a) = F_{2i,j-1}(a) + S_{i,j-1}(a)q_{2ij}(a), \tag{14b}$$

$$S_{ij}(a) = S_{i,j-1}(a)\exp\{-h_{Tij}(a)\}. \tag{14c}$$

This procedure estimates cumulative incidence from cause-specific models. It is distinct from fitting a Fine-Gray subdistribution hazard model (Fine & Gray, 1999). A cause-specific hazard ratio, a subdistribution hazard ratio, and a cumulative-incidence risk difference are different estimands and should not be described as if they were interchangeable.

This construction treats $-\log(1 - p_k)$ as a latent cause-specific integrated-hazard contribution. It is a grouped-time approximation that becomes more accurate as intervals become shorter and event probabilities become smaller; it is not an identity for arbitrary mutually exclusive cause probabilities. With coarse intervals, analysts should examine sensitivity to a finer time scale and compare the results with a direct weighted Aalen-Johansen estimator or an appropriate multinomial discrete-time model.

Figure 3 contrasts model-based standardization with the direct weighted Aalen-Johansen route.

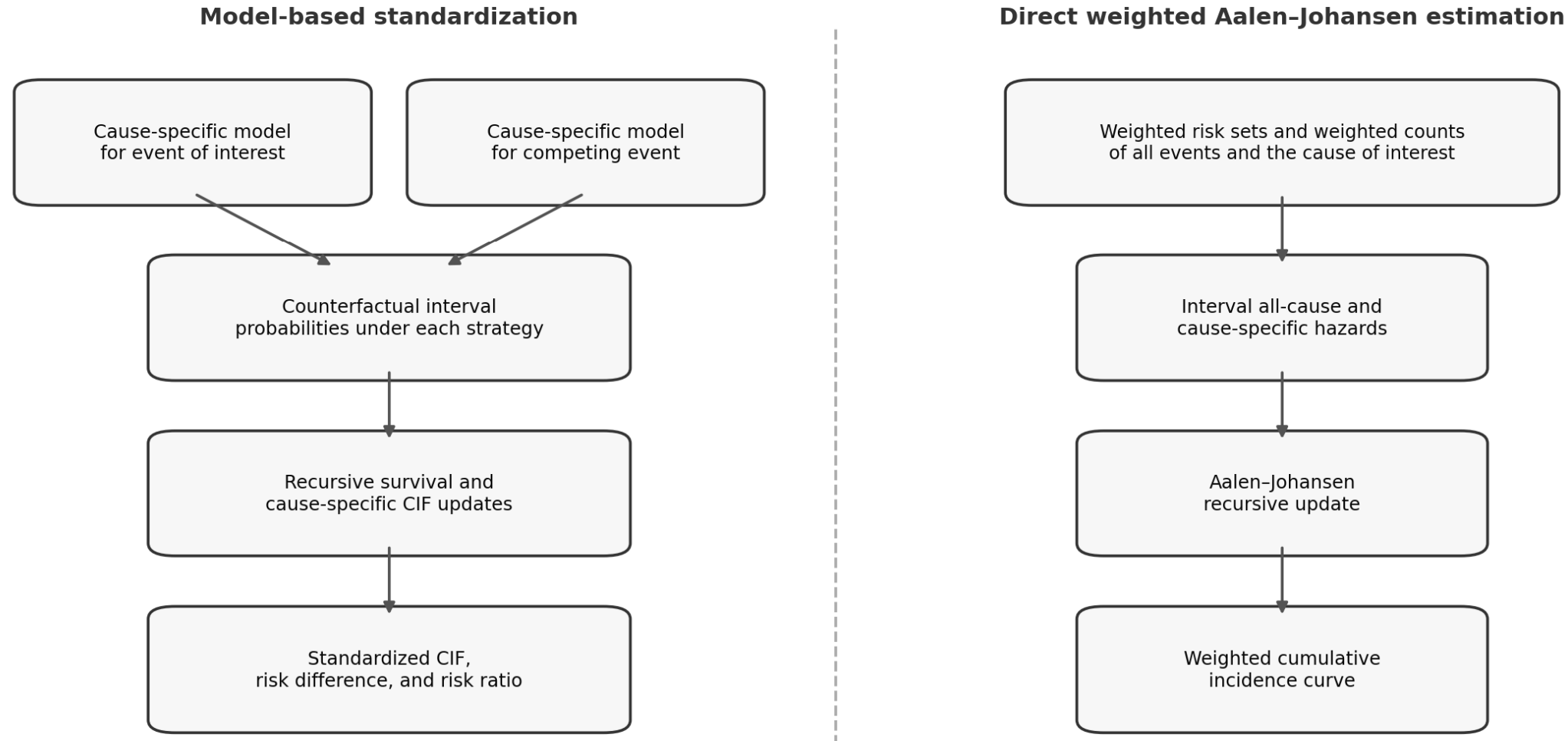

**Figure 3.** Two routes to cumulative incidence in TTE. Model-based standardization combines counterfactual probabilities from two cause-specific outcome models. The direct weighted Aalen-Johansen estimator updates cumulative incidence from weighted risk sets and observed event counts.

### *4.8. Direct weighted Kaplan-Meier and Aalen-Johansen curves*

curve_tte(type = "km") estimates a weighted Kaplan-Meier curve directly from person-period risk sets. For treatment group $a$ in interval $j$, let $Y_{aj}$ be the sum of weights among rows at risk and $d_{aj}$ the weighted event count. The update is

$$S_{aj} = S_{a,j-1}\left(1 - \frac{d_{aj}}{Y_{aj}}\right), \qquad F_{aj} = 1 - S_{aj}. \tag{15}$$

For the weighted Aalen-Johansen estimator, let $d_{\text{all},aj}$ be the weighted count of all event types and $d_{k,aj}$ the weighted count for cause $k$. Then

$$F_{k,aj} = F_{k,a,j-1} + S_{a,j-1}\frac{d_{k,aj}}{Y_{aj}}, \tag{16}$$

$$S_{aj} = S_{a,j-1}\left(1 - \frac{d_{\text{all},aj}}{Y_{aj}}\right). \tag{17}$$

Curves are stored and plotted as right-continuous step functions. The direct curves use the observed weighted risk sets and do not require an outcome regression. They still depend on the estimated weights, censoring rules, and empirical support. Model-based and direct curves need not be identical. A large discrepancy should prompt examination of the time function, treatment-by-time effects, extreme weights, effective risk sets, and whether both estimators target the same population and censoring regime.

The analysis weight attached to interval $j$ must be predictable at the start of that interval; it must depend only on treatment and covariate information observed before the interval-specific event opportunity.

### *4.9. Individual-cluster bootstrap*

boot_tte() samples unique original individual identifiers with replacement. Every person-period row and every sequential-trial entry belonging to a sampled person are retained together. Repeated selections receive distinct bootstrap identifiers so that they remain separate clusters.

For complete uncertainty propagation, the statistic function supplied to boot_tte() should re-estimate baseline treatment models, longitudinal censoring or adherence models, combined weights, outcome models, and standardized or direct curve estimates within every bootstrap replicate. Resampling person-period rows independently destroys the longitudinal and repeated-trial dependence structure and is not appropriate.

The internal bootstrap option in curve_tte() resamples original individuals while treating the supplied weights as fixed. It is useful for rapid exploration, but a publication-quality interval for an analysis with estimated weights should generally use the full pipeline bootstrap. Percentile intervals are simple to interpret, although other bootstrap intervals or influence-function methods may be considered in advanced analyses.

## 5. A GENERAL R WORKFLOW

The commands below assume TTE version 1.1.2, the first CRAN release. The package is available from CRAN, and the public GitHub repository provides the worked-example scripts. The package can be installed as follows.

```
# Install once from CRAN
# install.packages("TTE")

library(TTE)
packageVersion("TTE")
options(width = 110)
```

A standard analysis proceeds in the following order:

1. Specify and document the target trial protocol.
2. Construct baseline and person-period datasets with explicit time ordering.
3. Run structural checks.
4. Estimate baseline treatment weights.
5. Assess measured baseline balance and treatment-weight support.
6. Estimate longitudinal censoring or adherence weights using histories available before each outcome interval.
7. Combine, truncate, and diagnose the final analysis weight.
8. Fit a weighted discrete-time outcome model with clustering by original individual.
9. Standardize the model to marginal survival, risk, or cumulative incidence.
10. Estimate a direct weighted Kaplan-Meier or Aalen-Johansen curve as a complementary analysis.
11. Re-estimate the full pipeline in an individual-cluster bootstrap when complete uncertainty propagation is required.

A generic structural check is:

```
check <- check_tte(
  long_data,
  id = id,
  trial = trial,
  time = time,
  event = event,
  treatment = A,
  weights = analysis_weight
)

print(check)
```

A generic sequential-trial expansion is:

```
expanded <- seqdesign_tte(
  source_data,
  id = id,
  calendar_time = month,
```

```
  eligible = eligible,
  treatment = treatment,
  outcome = outcome,
  censor = censor,
  max_follow = 60,
  induction = 1
)
```

At each eligible treatment-decision time, a new trial entry is created. The analyst must still ensure that baseline variables are evaluated relative to that trial-specific time zero and that future information is not copied backward.

## 6. WORKED EXAMPLE 1: SGLT2 INHIBITOR VERSUS DPP-4 INHIBITOR INITIATION

***Synthetic-data warning.*** *SGLT2_baseline and SGLT2 are fully synthetic. They contain no actual patient records. All numerical results in this section are demonstrations of the analysis and have no clinical interpretation.*

### *6.1. Emulated target trial and datasets*

The first example represents an active-comparator new-user target trial among adults aged 75 years or older with type 2 diabetes and was motivated by an applied target trial emulation in this population (Noma et al., 2026a). The strategies are initiation of a dipeptidyl peptidase-4 inhibitor (DPP-4i) and initiation of a sodium-glucose cotransporter 2 inhibitor (SGLT2i). The outcome is all-cause death during post-induction follow-up after a one-month induction period. In the bundled synthetic data, time = 0 denotes the first post-induction interval, and the reported 60-month horizon therefore refers to 60 months of post-induction follow-up. The example should be interpreted as a post-induction estimand and does not include deaths occurring during the induction month. Table 3 gives the operational protocol represented by the synthetic data.

**Table 3.** Target trial specification for the synthetic SGLT2i versus DPP-4i example

| Component | Specification |
|---|---|
| Objective | Compare the effect of initiating SGLT2i versus DPP-4i on all-cause death among older adults with type 2 diabetes |
| Eligibility | Age 75 years or older, type 2 diabetes, and eligible as a new user of either drug class |
| Strategies | Initiate DPP-4i (A = 0) or initiate SGLT2i (A = 1) |
| Time zero | Date of qualifying treatment initiation |
| Induction | One month; the analysis clock begins after this interval, and events during the induction month are not included |
| Follow-up | Up to 60 post-induction person-month intervals, death, loss to follow-up, or administrative end |
| Primary estimand | Post-induction ITT analogue of the initiation strategy |
| Secondary estimand | Post-induction PP effect under sustained compatibility with the baseline strategy |
| Analysis | Baseline IPTW, longitudinal IPCW, weighted complementary log-log model, standardization, and weighted Kaplan-Meier estimation |

The baseline dataset contains 700 trial entries: 497 DPP-4i initiators and 203 SGLT2i initiators. The long dataset contains 20,257 person-month rows.

```
data(SGLT2_baseline)
data(SGLT2)

sglt2_baseline <- SGLT2_baseline
sglt2_long <- SGLT2

dim(sglt2_baseline)
dim(sglt2_long)
table(sglt2_baseline$treatment)
with(sglt2_long, tapply(Y_death, treatment, sum))

check_sglt2 <- check_tte(
  sglt2_long,
  id = id,
  trial = trial,
  time = time,
  event = Y_death,
  treatment = A
)
print(check_sglt2)
```

SGLT2_baseline is used for the baseline treatment model and balance assessment. SGLT2 is used for longitudinal censoring models, outcome models, and weighted curves. The same id-trial key is used when a baseline weight is copied to every person-month row.

### *6.2. Estimate the baseline treatment weight and assess balance*

The denominator model includes baseline variables associated with treatment choice and outcome. Quadratic terms are included for age and HbA1c to illustrate nonlinear specification. Component-level truncation is disabled because the final product will be truncated after the censoring component is added.

```
wt_a_sglt2 <- est_wt(
  A ~ age + I(age^2) + female + bmi +
    hba1c + I(hba1c^2) + egfr + proteinuria +
    prior_heart_failure + prior_stroke +
    recent_hospitalization + trial_period,
  data = sglt2_baseline,
  type = "treatment",
  stabilize = TRUE,
  truncate = c(0, 1)
)

summary(wt_a_sglt2)

bal_sglt2 <- balance_wt(
  A ~ age + I(age^2) + female + bmi +
    hba1c + I(hba1c^2) + egfr + proteinuria +
    prior_heart_failure + prior_stroke +
    recent_hospitalization + trial_period,
  data = sglt2_baseline,
  weights = wt_a_sglt2
)

print(bal_sglt2)
plot(bal_sglt2)
```

In the fixed synthetic example, the maximum absolute SMD decreased from 0.611 before weighting to 0.045 after weighting. The treatment-weight ESS was 599.4 among 700 baseline entries. The first result indicates that the specified treatment model balanced the measured columns included in the diagnostic. The second indicates a moderate information loss due to unequal weights, but not a collapse of the pseudo-population. Neither result addresses unmeasured confounding.

The treatment weight is then mapped from the baseline dataset to the person-month dataset.

```
key_baseline <- paste(sglt2_baseline$id, sglt2_baseline$trial, sep = ":")
key_long <- paste(sglt2_long$id, sglt2_long$trial, sep = ":")

sglt2_long$w_a_est <- weights(
  wt_a_sglt2,
  which = "untruncated"
)[match(key_long, key_baseline)]

stopifnot(!anyNA(sglt2_long$w_a_est))
```

### *6.3. Estimate the loss-to-follow-up weight and construct the ITT weight*

stay_ltfu indicates continuation with respect to loss to follow-up through the next interval. The denominator model includes measured baseline predictors, treatment, calendar period, and a spline for follow-up time. The numerator is reduced to treatment, time, and calendar period. lag = 1 assigns the cumulative weight available at the start of each outcome interval.

```
wt_ltfu_sglt2 <- est_wt(
  stay_ltfu ~ A + splines::ns(time, df = 3) +
    age + female + bmi + hba1c + egfr + proteinuria +
    prior_heart_failure + prior_stroke +
    recent_hospitalization + trial_period,
  numerator = stay_ltfu ~
    A + splines::ns(time, df = 3) + trial_period,
  data = sglt2_long,
  type = "censoring",
  id = id,
  trial = trial,
  time = time,
  cumulative = TRUE,
  lag = 1,
  stabilize = TRUE,
  truncate = c(0, 1)
```

```
)

wt_itt_sglt2 <- combine_wt(
  sglt2_long$w_a_est,
  wt_ltfu_sglt2,
  truncate = c(0.01, 0.99),
  normalize = "none"
)

sglt2_long$w_itt_est <- weights(wt_itt_sglt2)

# Use the bundled simulation-reference ITT weight to reproduce
# the fixed numerical results reported below.
sglt2_long$w_itt_analysis <- sglt2_long$w_itt

diag_itt_sglt2 <- diagnose_wt(
  w_itt_est,
  data = sglt2_long,
  treatment = treatment,
  time = time,
  id = id,
  trial = trial
)
print(diag_itt_sglt2)
plot(diag_itt_sglt2, type = "weights")
plot(diag_itt_sglt2, type = "risk_set")
```

The weight distribution should be inspected before fitting the outcome model. Important questions are whether one strategy has systematically larger weights, whether the 1st and 99th percentile truncation changes many rows, and whether effective risk sets remain adequate near 60 months. A late horizon with a very small effective risk set should not be emphasized merely because the nominal number at risk is nonzero. A complete report should record the median, 99th percentile, maximum, proportion modified by truncation, and treatment-specific effective risk sets at 36 and 60 months.

### *6.4. Fit the weighted outcome model*

The ITT outcome model includes baseline strategy, a natural spline for follow-up time, and trial period. Under a grouped proportional-hazards interpretation, the coefficient for A estimates an interval-specific integrated hazard ratio.

```
fit_death_itt <- discsurvreg(
  Y_death ~ A + splines::ns(time, df = 3) + trial_period,
  data = sglt2_long,
  id = id,
  weights = w_itt_analysis,
  family = quasibinomial(link = "cloglog"),
  var_method = "standard"
)

print(fit_death_itt)
summary(fit_death_itt)
confint(fit_death_itt, parm = "A", eform = TRUE)
```

Using the bundled simulation-reference ITT weight, the interval-specific integrated hazard ratio was 0.662 (95% confidence interval 0.407 to 1.077). This means that the fitted integrated hazard under SGLT2i initiation was estimated to be about 34% lower than under DPP-4i initiation, conditional on the model form and weighting. The confidence interval included 1, so the synthetic example does not provide a precise relative-effect estimate. More importantly, the relative integrated-hazard measure alone does not state the cumulative probability of death by 60 months.

The estimated weights constructed in Sections 6.2-6.3 illustrate the applied workflow and are used for diagnostics. The fixed numerical results in Sections 6 and 7 use the bundled simulation-reference weights, which are explicitly selected in the code so that the manuscript values are exactly reproducible. Results obtained after re-estimating the weights will generally differ.

### *6.5. Standardize to absolute risks*

The fitted outcome model is standardized over the original baseline population under A = 0 and A = 1.

```
std_death_itt <- std_tte(
  fit_death_itt,
  data = sglt2_baseline,
  treatment = A,
  time = time,
  times = 0:59,
  values = c(0, 1),
  labels = c("DPP-4i", "SGLT2i")
```

```
)

summary(std_death_itt, horizon = 36)
summary(std_death_itt, horizon = 60)
plot(std_death_itt, measure = "risk")
```

At 60 months, the analysis using the bundled simulation-reference ITT weight gave standardized risks of 0.448 under DPP-4i initiation and 0.325 under SGLT2i initiation, for a risk difference of -0.123. In the synthetic population, this corresponds to 12.3 fewer deaths per 100 people over 60 months and a risk ratio of approximately 0.73. The reciprocal of the absolute risk difference is about 8, which can be described as a time-specific number needed to treat only if the causal assumptions and intervention definition are credible. Because the data are synthetic, these quantities are used solely to demonstrate interpretation.

### *6.6. Compare with a direct weighted Kaplan-Meier curve*

```
km_death_itt <- curve_tte(
  sglt2_long,
  time = time,
  event = Y_death,
  treatment = treatment,
  weights = w_itt_analysis,
  type = "km",
  id = id,
  trial = trial
)

summary(km_death_itt, time = 36)
summary(km_death_itt, time = 60)
plot(km_death_itt, measure = "risk")
```

Using the bundled simulation-reference ITT weight, the weighted Kaplan-Meier risks at 60 months were 0.408 for DPP-4i and 0.296 for SGLT2i, an absolute difference of -0.112. These direct estimates are close in direction and magnitude to the standardized risks but are not identical. The standardized analysis uses a fitted time function and averages model-based counterfactual predictions; the weighted Kaplan-Meier estimator updates observed weighted risk sets without an outcome regression. A difference of this size is not automatically a failure, but it should lead the analyst to compare the curves over time, inspect late effective risk sets, and assess the outcome model.

Table 4 summarizes the main demonstration outputs. The balance and ESS entries use the estimated treatment weight, whereas the outcome entries use the bundled simulation-reference combined weight selected above.

**Table 4.** Selected results from the fully synthetic SGLT2i versus DPP-4i example

| Quantity | Synthetic result | Interpretation |
|---|---|---|
| Maximum absolute SMD | 0.611 before weighting; 0.045 after weighting | Measured baseline columns were well balanced by the specified treatment weight |
| Treatment-weight ESS | 599.4 of 700 entries | Weight variability reduced effective information, but the baseline pseudo-population remained reasonably large |
| ITT integrated hazard ratio | 0.662 (95% CI 0.407 to 1.077) | Relative interval-specific integrated hazard was lower under SGLT2i, but the interval included no difference |
| Standardized 60-month risk | 0.448 DPP-4i; 0.325 SGLT2i | Model-based risk difference -0.123 and risk ratio about 0.73 |
| Weighted KM 60-month risk | 0.408 DPP-4i; 0.296 SGLT2i | Direct weighted risk difference -0.112 |

### *6.7. Per-protocol analysis*

For the PP estimand, intervals after strategy deviation are excluded from the PP risk set, and a joint continuation indicator combines remaining observed and adherent.

```
sglt2_pp <- subset(sglt2_long, pp_at_risk == 1)

sglt2_pp$stay_pp <- as.integer(
  sglt2_pp$stay_ltfu == 1 &
    sglt2_pp$stay_adherent == 1
)

wt_censor_pp_sglt2 <- est_wt(
  stay_pp ~ A + splines::ns(time, df = 3) +
    age + female + bmi + hba1c + egfr + proteinuria +
    prior_heart_failure + prior_stroke +
    recent_hospitalization + trial_period,
```

```
  numerator = stay_pp ~
    A + splines::ns(time, df = 3) + trial_period,
  data = sglt2_pp,
  type = "censoring",
  id = id,
  trial = trial,
  time = time,
  cumulative = TRUE,
  lag = 1,
  stabilize = TRUE,
  truncate = c(0, 1)
)

wt_pp_sglt2 <- combine_wt(
  sglt2_pp$w_a_est,
  wt_censor_pp_sglt2,
  truncate = c(0.01, 0.99),
  normalize = "none"
)
sglt2_pp$w_pp_est <- weights(wt_pp_sglt2)

# Use the bundled simulation-reference PP weight for fixed examples.
sglt2_pp$w_pp_analysis <- sglt2_pp$w_pp

fit_death_pp <- discsurvreg(
  Y_death ~ A + splines::ns(time, df = 3) + trial_period,
  data = sglt2_pp,
  id = id,
  weights = w_pp_analysis,
  family = quasibinomial(link = "cloglog"),
  var_method = "standard"
)

std_death_pp <- std_tte(
  fit_death_pp,
  data = sglt2_baseline,
  treatment = A,
  time = time,
  times = 0:59,
  values = c(0, 1),
  labels = c("DPP-4i", "SGLT2i")
)
```

The PP coefficient describes the relative interval-specific integrated hazard under sustained strategy compatibility, not merely the association among people who happened to remain adherent. Its causal interpretation requires measured predictors sufficient to control the artificial censoring process. The synthetic SGLT2 dataset contains limited updated clinical information; this part of the example therefore demonstrates computational structure more strongly than substantive adequacy. In an applied study, time-updated predictors of discontinuation, switching, frailty, renal function, hospitalization, and treatment tolerance may be essential.

A full PP analysis should repeat the same diagnostics used for ITT: the distribution of the joint censoring weight, effective risk sets, standardized risks, a direct weighted Kaplan-Meier curve, and sensitivity analyses for adherence definitions and truncation. ITT and PP results should be presented as answers to different questions rather than as competing estimates of a single parameter.

### *6.8. Full-pipeline bootstrap*

The cluster-robust interval from discsurvreg() conditions on the estimated weights. To propagate uncertainty from all estimated components, the statistic function supplied to boot_tte() should reconstruct the baseline treatment model, longitudinal censoring model, combined weight, outcome model, and standardized 60-month contrast within every bootstrap sample. The complete reproducibility script should record the original-individual cluster variable, replicate count, random seed, failed replicates, and interval method.

```
# full_pipeline_statistic(d) must re-estimate all weight and outcome models
# and return a named numeric vector, for example the 60-month risk difference.
bootstrap_result <- boot_tte(
  sglt2_long,
  id = id,
  statistic = full_pipeline_statistic,
  R = 2000,
  conf_level = 0.95,
  seed = 20260730
)
print(bootstrap_result)
```

## 7. WORKED EXAMPLE 2: SEQUENTIAL ARB VERSUS CCB TRIALS WITH COMPETING DEATH

***Synthetic-data warning.*** *ARB_baseline and ARB are fully synthetic. The example is motivated by a clinical target trial emulation, but no record from that study is included and the numerical results have no clinical interpretation.*

### *7.1. Emulated target trial and repeated trial entries*

The second example represents sequentially nested new-user trials comparing angiotensin receptor blocker (ARB) and calcium channel blocker (CCB) strategies among people with chronic kidney disease and was motivated by an applied target trial emulation in this population (Noma et al., 2026b). Heart-failure hospitalization is the event of interest and death is a competing event. The baseline data contain 900 person-trial entries contributed by 750 original individuals; some people enter more than one trial. The long data contain 24,063 person-month rows. As in Example 1, time = 0 indexes the first post-induction interval, and the reported 60-month results refer to 60 months of post-induction follow-up.

**Table 5.** Target trial specification for the synthetic ARB versus CCB example

| Component | Specification |
|---|---|
| Objective | Compare ARB and CCB initiation strategies for heart-failure hospitalization in chronic kidney disease |
| Eligibility | Evaluated repeatedly at prespecified calendar times; an eligible person may enter more than one trial |
| Strategies | Initiate CCB (A = 0) or initiate ARB (A = 1) |
| Time zero | Each eligible treatment-decision time |
| Induction | One month; time = 0 denotes the first post-induction interval |
| Follow-up | Up to 60 post-induction person-month intervals, heart failure, death, loss to follow-up, or administrative end |
| Event of interest | Heart-failure hospitalization (event_code = 1) |
| Competing event | Death (event_code = 2) |
| Estimands | Post-induction ITT initiation effect and PP sustained-strategy effect, averaged over eligible person-trial entry occasions |
| Analysis | Baseline and longitudinal weighting, separate cause-specific models, standardized CIF, and weighted Aalen-Johansen estimation |

```
data(ARB_baseline)
data(ARB)

arb_baseline <- ARB_baseline
arb_long <- ARB

dim(arb_baseline)
dim(arb_long)
length(unique(arb_baseline$id))
nrow(arb_baseline)
```

Because trial entries repeat, baseline-to-long matching must use both id and trial. However, covariance estimation and bootstrap use the original id, because repeated trials from the same person are statistically dependent.

```
key_baseline <- paste(arb_baseline$id, arb_baseline$trial, sep = ":")
key_long <- paste(arb_long$id, arb_long$trial, sep = ":")
```

The same design can be constructed from an underlying longitudinal source table with seqdesign_tte(). Eligibility and treatment variables must be defined at each calendar decision time before the expansion is called.

### *7.2. Estimate and diagnose the ITT weights*

The treatment and loss-to-follow-up components follow the same logic as in Example 1. The ARB data additionally contain systolic blood pressure, diastolic blood pressure, and estimated glomerular filtration rate updated over follow-up. An applied denominator model for continuation can therefore include current or lagged values, provided that they are measured before the outcome interval being weighted.

```
wt_a_arb <- est_wt(
  A ~ age + I(age^2) + female + bmi + sbp + dbp + egfr +
    proteinuria + diabetes + prior_heart_failure + prior_stroke +
    trial_period,
```

```
  data = arb_baseline,
  type = "treatment",
  stabilize = TRUE,
  truncate = c(0, 1)
)

arb_long$w_a_est <- weights(
  wt_a_arb,
  which = "untruncated"
)[match(key_long, key_baseline)]

wt_ltfu_arb <- est_wt(
  stay_ltfu ~ A + splines::ns(time, df = 3) +
    age + female + bmi + sbp + dbp + egfr +
    proteinuria + diabetes + prior_heart_failure + prior_stroke +
    trial_period,
  numerator = stay_ltfu ~
    A + splines::ns(time, df = 3) + trial_period,
  data = arb_long,
  type = "censoring",
  id = id,
  trial = trial,
  time = time,
  cumulative = TRUE,
  lag = 1,
  stabilize = TRUE,
  truncate = c(0, 1)
)

wt_itt_arb <- combine_wt(
  arb_long$w_a_est,
  wt_ltfu_arb,
  truncate = c(0.01, 0.99),
  normalize = "none"
)

arb_long$w_itt_est <- weights(wt_itt_arb)

# Use the bundled simulation-reference ITT weight to reproduce
# the fixed numerical results reported below.
arb_long$w_itt_analysis <- arb_long$w_itt
```

In a sequential-trial analysis, diagnostics should be examined at three levels: baseline balance across person-trial entries, the overall and strategy-specific distribution of longitudinal weights, and effective risk sets over follow-up. The number of trial entries must not be mistaken for the number of independent people. The final report should also give the median, 99th percentile, maximum, proportion modified by truncation, and strategy-specific effective risk sets at 36 and 60 months.

### *7.3. Fit separate cause-specific outcome models*

One weighted model is fitted for heart-failure hospitalization and a second for the competing event of death. Under the grouped proportional-hazards interpretation, each coefficient for A is a cause-specific interval-specific integrated hazard ratio.

```
fit_hf_itt <- discsurvreg(
  Y_hf ~ A + splines::ns(time, df = 3) + trial_period,
  data = arb_long,
  id = id,
  weights = w_itt_analysis,
  family = quasibinomial(link = "cloglog"),
  var_method = "standard"
)

fit_death_competing <- discsurvreg(
  Y_death ~ A + splines::ns(time, df = 3) + trial_period,
  data = arb_long,
  id = id,
  weights = w_itt_analysis,
  family = quasibinomial(link = "cloglog"),
  var_method = "standard"
)

confint(fit_hf_itt, parm = "A", eform = TRUE)
confint(fit_death_competing, parm = "A", eform = TRUE)
```

Using the bundled simulation-reference ITT weight, the cause-specific integrated hazard ratio was 0.829 (95% CI 0.592 to 1.161) for heart-failure hospitalization and 0.539 (95% CI 0.346 to 0.840) for death. The confidence interval for heart-failure hospitalization included 1, whereas that for death did not. These are cause-specific rates among person-trials that are alive and free of both heart-failure hospitalization and competing death at the start of the interval. They do not directly give the 60-month probability of heart-failure hospitalization.

### 7.4. Standardize to a cumulative incidence function

Both cause-specific models are supplied to std_tte(). Because arb_baseline contains one row per eligible person-trial entry, the standardized estimand averages over the empirical distribution of eligible treatment-decision occasions rather than over unique individuals. Individuals eligible on multiple occasions therefore contribute multiple target-population rows. A unique-person target would require a separate one-row-per-person target dataset or person-level target weights that equalize each individual's total contribution.

```
std_hf_itt <- std_tte(
  fit_hf_itt,
  data = arb_baseline,
  treatment = A,
  time = time,
  times = 0:59,
  competing_fit = fit_death_competing,
  values = c(0, 1),
  labels = c("CCB", "ARB")
)

summary(std_hf_itt, horizon = 36)
summary(std_hf_itt, horizon = 60)
plot(std_hf_itt, measure = "risk")
```

Using the bundled simulation-reference ITT weight, the 60-month standardized heart-failure CIF was 0.350 under CCB and 0.330 under ARB, for a risk difference of -0.020. The relative cause-specific integrated-hazard estimate suggested a 17% lower heart-failure hazard, yet the absolute cumulative-incidence difference was only 2 percentage points. This is not contradictory. Cumulative incidence depends on both the heart-failure hazard and the competing death hazard. A treatment strategy that changes survival can alter the number of people who remain available to experience heart failure.

### 7.5. Direct weighted Aalen-Johansen estimation

```
aj_hf_itt <- curve_tte(
  arb_long,
  time = time,
  event = event_code,
  treatment = treatment,
  weights = w_itt_analysis,
  type = "aj",
  cause = 1,
  id = id,
  trial = trial
)

summary(aj_hf_itt, time = 36)
summary(aj_hf_itt, time = 60)
plot(aj_hf_itt, measure = "risk")
```

Using the bundled simulation-reference ITT weight, the weighted Aalen-Johansen CIFs at 60 months were 0.349 under CCB and 0.298 under ARB, a difference of -0.051. The CCB estimate is almost identical to the standardized value, whereas the ARB estimate is lower. The discrepancy should not be hidden. It directs attention to the ARB outcome-model form, treatment-by-time behavior, late effective risk sets, model smoothing, and the grouped-time latent-hazard approximation used to combine the two cause-specific models. In a real analysis, one would compare the entire curves and repeat the standardization under alternative time specifications.

Using one minus a weighted Kaplan-Meier estimator for heart-failure hospitalization would be incorrect because death is a competing event. The relevant direct estimator is Aalen-Johansen.

### 7.6. Per-protocol analysis and interpretation

The PP analysis restricts to intervals compatible with the assigned strategy, estimates a joint censoring weight for remaining observed and adherent, and then repeats both cause-specific models and both routes to cumulative incidence.

```
arb_pp <- subset(arb_long, pp_at_risk == 1)

arb_pp$stay_pp <- as.integer(
  arb_pp$stay_ltfu == 1 &
    arb_pp$stay_adherent == 1
)
```

```
wt_censor_pp_arb <- est_wt(
  stay_pp ~ A + splines::ns(time, df = 3) +
    age + female + bmi + sbp + dbp + egfr +
    proteinuria + diabetes +
    prior_heart_failure + prior_stroke +
    trial_period,
  numerator = stay_pp ~
    A + splines::ns(time, df = 3) + trial_period,
  data = arb_pp,
  type = "censoring",
  id = id,
  trial = trial,
  time = time,
  cumulative = TRUE,
  lag = 1,
  stabilize = TRUE,
  truncate = c(0, 1)
)

wt_pp_arb <- combine_wt(
  arb_pp$w_a_est,
  wt_censor_pp_arb,
  truncate = c(0.01, 0.99),
  normalize = "none"
)
arb_pp$w_pp_est <- weights(wt_pp_arb)

diag_pp_arb <- diagnose_wt(
  w_pp_est,
  data = arb_pp,
  treatment = treatment,
  time = time,
  id = id,
  trial = trial
)
print(diag_pp_arb)
plot(diag_pp_arb, type = "weights")
plot(diag_pp_arb, type = "risk_set")

# Use the bundled simulation-reference PP weight for fixed results.
arb_pp$w_pp_analysis <- arb_pp$w_pp

fit_hf_pp <- discsurvreg(
  Y_hf ~ A + splines::ns(time, df = 3) + trial_period,
  data = arb_pp,
  id = id,
  weights = w_pp_analysis,
  family = quasibinomial(link = "cloglog"),
  var_method = "standard"
)

fit_death_pp <- discsurvreg(
  Y_death ~ A + splines::ns(time, df = 3) + trial_period,
  data = arb_pp,
  id = id,
  weights = w_pp_analysis,
  family = quasibinomial(link = "cloglog"),
  var_method = "standard"
)

std_hf_pp <- std_tte(
  fit_hf_pp,
  data = arb_baseline,
  treatment = A,
  time = time,
  times = 0:59,
  competing_fit = fit_death_pp,
  values = c(0, 1),
  labels = c("CCB", "ARB")
)
summary(std_hf_pp, horizon = 36)
summary(std_hf_pp, horizon = 60)
plot(std_hf_pp, measure = "risk")

aj_hf_pp <- curve_tte(
  arb_pp,
  time = time,
  event = event_code,
  treatment = treatment,
  weights = w_pp_analysis,
  type = "aj",
  cause = 1,
```

```
  id = id,
  trial = trial
)
summary(aj_hf_pp, time = 36)
summary(aj_hf_pp, time = 60)
plot(aj_hf_pp, measure = "risk")
```

Using the bundled simulation-reference PP weight, the cause-specific integrated hazard ratio for heart-failure hospitalization was 0.516 (95% CI 0.275 to 0.965). The standardized 60-month heart-failure CIF was 0.250 under sustained CCB and 0.144 under sustained ARB, for a risk difference of -0.107. In the synthetic example, the PP contrast is therefore substantially larger than the ITT contrast. This should be interpreted as a difference between initiation and sustained-strategy estimands, not as proof that one analysis is less biased. The PP result also rests on stronger assumptions about time-varying exchangeability and adherence positivity.

Table 6 summarizes the fixed simulation-reference outputs.

**Table 6.** Selected results from the fully synthetic ARB versus CCB example

| Analysis | Relative-effect model | 60-month model-based absolute effect | 60-month direct weighted curve |
|---|---|---|---|
| ITT, heart failure with competing death | HF cause-specific integrated HR 0.829 (95% CI 0.592 to 1.161); death integrated HR 0.539 (0.346 to 0.840) | HF CIF 0.350 CCB and 0.330 ARB; RD -0.020 | Weighted AJ CIF 0.349 CCB and 0.298 ARB; RD -0.051 |
| PP, heart failure with competing death | HF cause-specific integrated HR 0.516 (95% CI 0.275 to 0.965) | HF CIF 0.250 sustained CCB and 0.144 sustained ARB; RD -0.107 | PP weighted AJ curve obtained from the displayed PP risk set and bundled simulation-reference PP weights |

The example demonstrates why a complete competing-risk analysis should report at least one relative cause-specific measure and a clinically interpretable cumulative-incidence contrast. Reporting only the heart-failure hazard ratio would omit the influence of competing death; reporting only the CIF would make it difficult to understand which cause-specific process generated the result.

## 8. DIAGNOSTICS, SENSITIVITY ANALYSES, AND REPORTING

### *8.1. A practical diagnostic sequence*

Diagnostics should be performed before the treatment-effect result is treated as final. Table 7 gives a practical sequence.

**Table 7.** Diagnostic findings and recommended responses

| Finding | Possible explanation | Recommended response |
|---|---|---|
| Large baseline SMD after weighting | Treatment model misspecification, missing interaction/nonlinearity, poor support | Revise scientifically justified functional forms; inspect overlap; consider restriction or a different target population |
| Extreme treatment weights concentrated in one strategy | Near-positivity violation or highly deterministic treatment choice | Identify covariate histories causing extremes; reconsider eligibility/comparator; report sensitivity to truncation |
| Acceptable overall ESS but very small late effective risk sets | Attrition or cumulative weighting concentrates information | Shorten the primary horizon, report risk-set diagnostics, or improve censoring models if justified |
| Weight distribution changes sharply after a particular time | Measurement or model discontinuity, calendar effects, sparse histories | Inspect interval-specific continuation probabilities and data coding around that time |
| Standardized and direct weighted curves diverge | Outcome-model misspecification, sparse risk sets, nonproportional effects | Compare time functions and interactions; inspect both curves and effective risk sets; use bootstrap |
| PP estimate much larger than ITT estimate | Different estimands, treatment changes, artificial-censoring model, adherence positivity | Describe both causal questions; diagnose PP weights; examine strategy-deviation definitions |
| Competing-risk CIF behaves unexpectedly | Strong effect on competing event, event-code errors, incorrect use of one minus KM | Inspect both cause-specific models and event coding; use standardized CIF or Aalen-Johansen |

A diagnostic threshold should not be applied mechanically. For example, an SMD of 0.09 is not automatically acceptable if it concerns a dominant prognostic variable and other diagnostics suggest poor overlap. Conversely, a small residual imbalance can be addressed by including the covariate in the outcome model, provided that the analysis plan and interpretation remain transparent.

### *8.2. Sensitivity analyses*

At minimum, an applied analysis should consider sensitivity to the following choices:

- treatment and censoring model functional forms, including nonlinear terms and interactions;
- alternative reasonable weight truncation limits, such as no truncation, 0.5th/99.5th, and 1st/99th percentiles;
- follow-up-time representation and treatment-by-time interactions;
- primary analysis horizon, especially when late effective risk sets are small;
- definitions of treatment discontinuation, switching, add-on therapy, and allowable gaps for PP analyses;
- induction or lag choices that follow from the measurement process;
- target population used for standardization;
- handling of missing baseline and time-varying covariates;
- competing-event definitions and cause coding; and
- full-pipeline bootstrap versus intervals that treat estimated weights as fixed.

Sensitivity analyses should be linked to plausible alternative versions of the target trial or analysis assumptions. A large menu of arbitrary models is less informative than a small set of clearly motivated alternatives.

### *8.3. Recommended reporting and reproducibility archive*

A study using TTE should report the complete target trial protocol, operational definitions, baseline and person-period data structures, numerator and denominator weight models, temporal ordering and lag, truncation and normalization rules, balance and support diagnostics, effective risk sets, outcome-model link and time function, target population for standardization, relative and absolute effects, competing-risk handling, bootstrap cluster definition, and sensitivity analyses. The TARGET Statement provides a structured reporting framework (Cashin et al., 2025).

The reproducibility archive should contain:

- the exact TTE and R versions;
- code that constructs baseline and person-period data;
- the synthetic-data generation code, parameter settings, and random seeds;
- a master script that regenerates all numerical results and tables in the article;
- a dependency lock file or equivalent record of the package environment;
- all numerator and denominator formulas;
- weight truncation and normalization settings;
- balance, weight, ESS, and effective-risk-set outputs;
- outcome-model formulas, family, link, and covariance option;
- the standardization population and any `target_weights`;
- competing-event definitions and event codes;
- bootstrap statistic function, cluster variable, replicate count, and random seed;
- complete `sessionInfo()` output; and
- a human-readable protocol that permits an independent analyst to reconstruct time zero and every censoring rule.

A version-specific software citation can be obtained with `citation("TTE")`. Because defaults and returned objects can evolve, the version should be stored with the analysis rather than inferred later from a repository’s current state.

The release accompanying the preprint should be tagged and immutable; a DOI-backed archive is preferable for long-term reproducibility.

## 9. SCOPE, RELATIONSHIP TO OTHER SOFTWARE, AND LIMITATIONS

Several packages with broader or more specialized target trial functionality have been developed. TrialEmulation provides a mature framework for sequences of emulated trials and includes scalable storage and sampling options (Su et al., 2024). SEQTaRget supports static and dynamic strategies with time-varying treatments and confounders (O’Dea et al., 2026). TTE focuses on a compact static-strategy workflow with explicit intermediate quantities, direct weighted Kaplan-Meier and Aalen-Johansen estimation, cause-specific competing-risk standardization, a unified elapsed-time convention, and inference clustered or resampled at the original-individual level.

Version 1.1.2 does not provide a general implementation of dynamic treatment regimes, complex cloning and grace-period strategies, continuous treatments, targeted maximum likelihood estimation, cross-fitted machine-learning nuisance models, integrated multiple imputation, or database-backed processing for extremely large datasets. `balance_wt()` focuses on binary comparisons, although `est_wt()` can estimate multinomial treatment probabilities. The default nuisance and outcome models are parametric and can be misspecified. The internal curve bootstrap treats supplied weights as fixed. Full uncertainty propagation requires `boot_tte()` with re-estimation of all components. std_tte() does not simulate treatment-dependent post-baseline covariate trajectories.

The package also cannot determine whether a target trial is clinically meaningful. It cannot resolve poorly measured treatment, unobserved treatment versions, outcome misclassification, inappropriate carry-forward of covariates, lack of overlap, or unmeasured confounding. A transparent implementation can make these problems easier to detect and discuss, but not make them disappear.

## 10. DISCUSSION AND CONCLUSION

Target trial emulation is sometimes described as a statistical method, but its most important contribution is the disciplined connection between a causal question, a protocol, a longitudinal data structure, and an analysis. The weighting model, outcome regression, and standardization are meaningful only after the strategies, time zero, follow-up, censoring, and estimand have been defined. TTE reflects that order by beginning with constructed trial data and exposing each subsequent step.

The package’s modularity is useful in practical work. An analyst can inspect treatment probabilities before weighting, distinguish interval-specific from cumulative censoring factors, evaluate what truncation changed, compare nominal with effective risk sets, and compare model-based standardized curves with direct weighted curves. A collaborator or reviewer can identify which component generated a result rather than receiving only a final hazard ratio from a black box.

The two synthetic examples illustrate complementary situations. The SGLT2i example shows an active-comparator new-user design, baseline and longitudinal weighting, ITT and PP estimands, a weighted complementary log-log model, standardized absolute risk, and a weighted Kaplan-Meier curve. The ARB example adds repeated trial entries and competing death, demonstrating the distinction between cause-specific hazards and cumulative incidence and the use of both standardized and direct Aalen-Johansen estimates.

The examples also show why interpretation cannot end with a single coefficient. A hazard ratio can coexist with a modest or large absolute risk difference depending on baseline risk, follow-up, and competing events. ITT and PP effects can differ because they compare different strategies. Standardized and direct curves can differ because one uses an outcome model and the other uses weighted risk sets. These differences are analytically useful: they identify where assumptions and empirical support deserve closer examination.

In conclusion, TTE provides a coherent R workflow for standard target trial emulations with longitudinal observational data. It links data checking and sequential-trial construction to inverse probability weighting, diagnostics, weighted discrete-time outcome models, standardization, competing-risk analysis, direct weighted curves, and cluster bootstrap at the original-individual level. Used with a defensible protocol and explicit causal assumptions, it can support transparent, reproducible, and practically interpretable comparative-effectiveness analyses.

# APPENDIX A. FUNCTION MAP AND SELECTED DEFAULTS

**Table A1**. Public functions, returned objects, and principal retained components

| Function | Primary task | Principal retained components |
|---|---|---|
| check_tte() | Structural checks | Report, validity flag, row count, original-person count, and person-trial count |
| seqdesign_tte() | Sequential-trial expansion | trial, time, baseline treatment, analysis outcome, risk-set indicators, and source-row mapping |
| est_wt() | Treatment, censoring, or adherence weights | Fitted numerator and denominator models, probability matrices, observed probabilities, interval factors, cumulative weights, untruncated and truncated weights |
| combine_wt() | Weight multiplication, truncation, normalization | Component matrix, raw product, final weight, and truncation limits |
| balance_wt() | Measured baseline balance | Unweighted and weighted means, SDs, SMDs, and maximum imbalance |
| diagnose_wt() | Weight and risk-set diagnostics | Overview, quantiles, group summaries, ESS, and effective risk sets |
| discsurvreg() | Weighted discrete-time model | GLM object, coefficients, cluster-robust covariance, analysis weights, and cluster identifiers |
| std_tte() | Model-based marginal effects | Strategy curves, contrasts, fitted models, and individual terminal predictions |
| curve_tte() | Weighted KM or AJ curve | Weighted risk-set and event summaries, survival/risk curves, and optional bootstrap values |
| boot_tte() | Individual-cluster bootstrap | Original estimate, replicate matrix, standard errors, and percentile intervals |

**Table A2**. Selected default arguments and their practical implications in TTE 1.1.2

| Function and default | Practical implication |
|---|---|
| est_wt(stabilize = TRUE) | Treatment weights use marginal numerator probabilities by default; longitudinal weights use an intercept-only numerator when no numerator formula is supplied |
| est_wt(truncate = c(0.01, 0.99)) | The component analysis weight is truncated after cumulative construction unless overridden; use c(0, 1) to disable truncation and report the selected rule explicitly |
| est_wt(cumulative = NULL) | Defaults to FALSE for treatment weights and TRUE for censoring/adherence weights |
| est_wt(lag = NULL) | Defaults to 0 for noncumulative weights and 1 for cumulative weights |
| combine_wt(normalize = "none") | No normalization after multiplication and truncation |
| balance_wt(absolute = TRUE) | Absolute SMDs are reported |
| discsurvreg(quasibinomial(link = "cloglog")) | Complementary log-log mean model is fitted by default |
| discsurvreg(var_method = "standard") | Individual-cluster HC0 sandwich covariance is used |
| std_tte(target_weights = NULL) | Target-population rows are averaged equally |
| curve_tte(bootstrap = 0) | Point estimates are returned without internal bootstrap intervals |
| boot_tte(R = 500, conf_level = 0.95) | 500 individual-cluster replicates and 95% percentile intervals are used |

## APPENDIX B. SYNTHETIC DATA DICTIONARIES

The four package datasets are independently simulated and contain no records from the motivating clinical studies. ps_oracle and the bundled weights are simulation reference quantities. Applied analyses must estimate weights from protocol-specific treatment, censoring, and adherence models.

**Table B1**. Variables in SGLT2_baseline and SGLT2

| Variable | Availability | Description |
|---|---|---|
| id | Baseline and long | Original individual identifier |
| trial | Baseline and long | Trial identifier; one trial entry per individual in this example |
| A | Baseline and long | 0 = DPP-4i, 1 = SGLT2i |
| treatment | Baseline and long | Display label for treatment strategy |
| age | Baseline and long | Age at trial baseline, years |
| female | Baseline and long | Indicator for female sex |
| bmi | Baseline and long | Body mass index, kg/m² |
| hba1c | Baseline and long | Glycated hemoglobin, percent |
| egfr | Baseline and long | Estimated glomerular filtration rate, mL/min/1.73 m² |
| proteinuria | Baseline and long | Indicator for proteinuria |
| prior_heart_failure | Baseline and long | Indicator for prior heart failure |
| prior_stroke | Baseline and long | Indicator for prior stroke |
| recent_hospitalization | Baseline and long | Indicator for a recent hospitalization |
| trial_period | Baseline and long | Calendar-period category at trial baseline |
| ps_oracle | Baseline only | Probability of SGLT2i initiation generated by the simulation treatment model |
| w_iptw | Baseline and long | Stabilized baseline reference treatment weight |
| time | Long only | Zero-based follow-up-month index |
| Y_death | Long only | Indicator of death during the interval |
| event_code | Long only | 0 = no event, 1 = death |
| stay_ltfu | Long only | Indicator of remaining observed through the next interval |
| stay_adherent | Long only | Indicator of strategy adherence through the next interval |
| pp_at_risk | Long only | Indicator that the trial entry remains in the PP risk set |
| w_itt | Long only | Bundled combined reference weight for the ITT analysis |
| w_pp | Long only | Bundled combined reference weight for the PP analysis |

**Table B2**. Variables in ARB_baseline and ARB

| Variable | Availability | Description |
|---|---|---|
| id | Baseline and long | Original individual identifier |
| trial | Baseline and long | Identifier for the sequential trial entry |
| A | Baseline and long | 0 = CCB, 1 = ARB |
| treatment | Baseline and long | Display label for treatment strategy |
| age | Baseline and long | Age at trial baseline, years |
| female | Baseline and long | Indicator for female sex |
| bmi | Baseline and long | Body mass index, kg/m² |
| sbp | Baseline and long | Systolic blood pressure, mm Hg; updated over follow-up in the long data |
| dbp | Baseline and long | Diastolic blood pressure, mm Hg; updated over follow-up in the long data |
| egfr | Baseline and long | eGFR, mL/min/1.73 m²; updated over follow-up in the long data |
| proteinuria | Baseline and long | Indicator for proteinuria |
| diabetes | Baseline and long | Indicator for diabetes |
| prior_heart_failure | Baseline and long | Indicator for prior heart failure |
| prior_stroke | Baseline and long | Indicator for prior stroke |
| trial_period | Baseline and long | Calendar-period category at trial baseline |
| ps_oracle | Baseline only | Probability of ARB initiation generated by the simulation treatment model |
| w_iptw | Baseline and long | Stabilized baseline reference treatment weight |
| time | Long only | Zero-based follow-up-month index |

| Variable | Availability | Description |
|---|---|---|
| Y_hf | Long only | Indicator of heart-failure hospitalization during the interval |
| Y_death | Long only | Indicator of death during the interval |
| event_code | Long only | 0 = no event, 1 = heart-failure hospitalization, 2 = death |
| stay_ltfu | Long only | Indicator of remaining observed through the next interval |
| stay_adherent | Long only | Indicator of strategy adherence through the next interval |
| pp_at_risk | Long only | Indicator that the person-trial remains in the PP risk set |
| w_itt | Long only | Bundled combined reference weight for the ITT analysis |
| w_pp | Long only | Bundled combined reference weight for the PP analysis |

**Funding.** This study was supported by the Japan Society for the Promotion of Science (JP24K21306, JP23K11931 and JP22H03554). The funders had no role in the design or implementation of the software, generation or analysis of the synthetic data, preparation of the manuscript, or decision to disseminate the work.

**Data availability.** All data used in the examples are fully synthetic and distributed with the TTE package. No actual patient-level records from the motivating studies are included.

**Code availability.** TTE version 1.1.2 is available from CRAN at https://doi.org/10.32614/CRAN.package.TTE. The public GitHub repository at https://github.com/nomahi/TTE provides the worked-example scripts.